\documentclass[conference]{IEEEtran}
\usepackage{dcolumn}  

\usepackage{amsmath} 
\ifCLASSINFOpdf
\usepackage[pdftex]{graphicx}
\graphicspath{{../pdf/}{../jpeg/}{../png/}}
 
\DeclareGraphicsExtensions{.pdf,.jpeg,.png}
\else
}
\fi
\begin{document}
\title{Better User Recommendations using Enhancing Software Development Process  Repository}
\author{\IEEEauthorblockN{Ziaur Rahman, Md. Kamrul Hasan}
	\IEEEauthorblockA{Department Computer Science and Engineering
       \\  Systems and Software Lab (SSL)
		\\ Islamic University of Technology (IUT),
		Bangladesh\\
		Email: {\{zia, hasank\}@iut-dhaka.com}
		}}

\maketitle

\begin{abstract}
%\boldmath
Reusing previously completed software repository to enhance the development process is a common phenomenon. If developers get suggestions from the existing projects they might be benefited a lot what they eventually expect while coding. The strategies available in this field have been rapidly changing day by day. There are a number of efforts that have been focusing on mining process and constructing repository. Some of them have emphasized on the web based code searching while others have integrated web based code searching in their customized tool. But web based approaches have inefficiency especially in building repository on which they apply mining technologies. To search the code snippets in response to the user query we need an enriched repository with a better representation and abstraction. To ensure that repository before mining process we have developed a concept based on Enhancing Software Development Process (ESDP). In ESDP approach multiple sources of codes from both online and offline storages are considered to construct the central repository with XML representation and applied mining techniques in the client side. The respective evaluation shows that ESDP approach works much better in response time and performance than many other existing approaches available today. 
\end{abstract}

\section{Introduction}

%\em

% no \IEEEPARstart

The application of data mining has great advantages and potentials in developing software. A software developer can be guided with the knowledge extracted from the previously completed software projects and artifacts. The source files, documentations and associated files can be a rich source for building repository. To extract knowledge, snippets or any other recommendation a number of similar projects should be mined following the proper strategy. Before applying the mining algorithm repository building process should be considered with the highest priority. The recommendation will be more accurate and relevant if the repository on which the data mining technique is applied is more updated and enriched. There are some approaches available where the repository building is avoided \cite{9} or any other web repository \cite{4}, \cite{5} is used before the mining process. If recommendation system or tool only depends on web based code searching system then a lot of issues are needed to resolve. The storage of the web search tool on which we search is not well documented and structured. In most cases the abstraction of the sources are not well represented. The searching algorithms that the code search systems use are almost same with that the general search engine applies. The general search engine often deals with the extra large files that are mostly not convenient to the software code seekers. It is also difficult for the developer to choose which one is more relevant to him. The general searching algorithm like Pigeon Rank, Spider and Crawler that they apply is not good enough to fulfill the developer's needs. The search result is normally indexed or ranked as per the title, heading and meta-information of the source files irrespective what actually it has inside. Web based searching has server dependency. Even in the distributed system it has multiple server dependency that often cause hazards and consume valuable time of the developers. 

It is seen in the existing approaches \cite{5}, \cite{9} that a code search engine is used to get the desired item following a search query. In some cases, searching happened before instant mining that we call post-mining strategy. In doing so search dimension gets excessively larger. That is a clear hindrance against finding exact match. But if the search domain is fixed only in the repository of the code search engine then it will obviously keep us away from getting the exact match. Rapidly searching online has some drawbacks itself. Sometimes it takes longer time due to request and response latency of the server.This inconveniently kills valuable time of the programmers. However, security authentication and connection issues are also vital issue. It gets threatened when it happens over the Internet. How much a developer is likely to get connected with an unknown server while he is on the version control system that does matter?

Considering these issues a concept of a system that we call Enhancing Software Development Process (ESDP) with an enriched repository and better code abstraction can help the developers in the client.The paper shows that the development performance is highly influenced by using the Enhancing Software Development Process (ESDP) repository.

\section{Background}
The application of data mining technique has great advantages and potentials in developing software. Software developer often needs searching existing project repositories. Using the code searching tool is one of the existing approaches that can guide developer by providing related code snippet and patterns. There are a number of efforts found that influences the software development process.PR Miner \cite{1}, Taxonomy approach of Mining Repositories \cite{2}, Perracotta \cite{3}, MAPO \cite{4},\cite{5}, XSnippet \cite{6}, Mining API Pattern \cite{7}, PARSEWeb \cite{8}, MAC \cite{9}, Scenario Based API Recommendation System \cite{10} are some of the popular efforts in this area. Mining API Usages from Open Source Repositories (MAPO) \cite{4}, \cite{5} is one of the earliest and MAC \cite{9} is one of the recent effort to mine API usage pattern. Here we have explained different existing approaches that are widely used in mining software repository (MSR). 

PR-Miner \cite{1} extracts programming rules in general form and propose algorithms to detect rule violations. Taxonomy approach \cite{2} deals with the software artifacts or temporal information where a demonstrated and expressive taxonomy is derived from the analysis of this literature and presents the work via four dimensions. Perracotta \cite{3} works with the scaling dynamic inference techniques and also deals with the large programs through the imperfect traces along with approximate inference algorithm.

 MAPO \cite{4}, \cite{5} is able to identify call patterns from the API usages of an existing project. It works on a query that describes a method, class,  package for a particular API. MAC \cite{9} mines API code snippets for code reuse. It forms a transaction database. Then a pattern database is formed from the transaction database. According to the initial statement MAC is able to predict useful related API code snippets according to the initial statement. Thus it guides developers through related API usage patterns by evaluating the support, confidence and rank list of the frequent item.

Strathcona \cite{11} gives a number of relevant snippets by matching the structure of the code under development with the snippets belongs to the repository. CodeBroker \cite{12} is mostly similar tool to Strathcona. It automatically searches the repository by using comments provided by the developer.  CodeFinder \cite{13} uses a query browser to help the developer to construct the queries that can be sent to the repository.  XSnippet \cite{6} was developed by Tansalarak and Claypool. They extend Prospector and add additional queries, ranking heuristics and mining algorithms to query a code snippet repository for the relevant snippet at hand. PARSEWeb \cite{8} developed by Thummalapenta and Xie used Google code search for collecting relevant code snippets and mines the returned code snippets to find the solution. 

Saul proposed an approach \cite{17} to find API methods that are closely related to a query API method of interest by discovering API methods.Then it shares a caller or a callee with the query API method. 

 Another attempt is GrouMiner \cite{18}, \cite{19} a novel graph-based approach for mining the usage patterns of one or multiple objects. GrouMiner approach includes a graph-based representation for the multiple object usages, a pattern mining algorithm and an anomaly detection technique that are efficient, accurate and resilient to the software changes. The tool that automatically builds queries to send to the repository is the Hipikat tool. Hipikat \cite{20} creates links between different sources of information in a project, including source files, cvs commits, bug reports, newsgroup postings, and web articles. 

In most of the related and existing works they either have build a customized repository using single source of projects or have used different code search system away from their framework. ESDP approach has its own repository in the client side that is built using different sources of projects and files. The difference is the use of particular source extraction technique before applying mining technique. We have considered our repository in the client side to avoid request-response latency problems to accelerate development process.

\begin{figure}[h]
\centering
\includegraphics[height=4.5in]{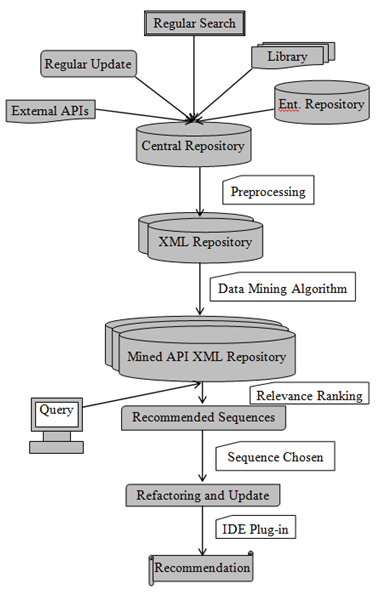}
\caption{Enhancing Software Development Process (ESDP) Framework.}
\end{figure}
\section{Proposed Idea}

ESDP has enriched and updated repository building strategy. We have applied ESDP repository to recommend the related snippets in response to the user queries. We have initially implemented the system as Integrated Development Environment (IDE) plug-in and consequently we have been developing our own working environment.  In our system we have three heuristics that we respectively call as Building Repository, Source Abstraction for Data Mining and Searching and Recommendation Heuristics. All of the three steps belongs the ESDP framework.  The framework is shown in the above Figure 1. In our ESDP system we have designed the system without server dependency. Even for storing the mined sources we have applied XML based system \cite{21} in the client system. That is why it is able to handle multiple search queries simultaneously without any concurrency problem. ESDP repository is kept away from the server failure problem. The following steps will explain the three heuristics of the framework by turn. 

\subsection{Building Repository}
Repository building is important in mining software repository. If the repository is more updated and enriched then the mined system will be able to recommend relevant suggestions and patterns. There are a number of issues should be considered before building a repository. Two of these are very fundamental that are listed below. 

\begin{enumerate}
\item
What are the sources of codes that are used to build the repository?
\item
 How often the repository is updated?
\end{enumerate}

If the sources are limited then the mined repository will be comparatively light weighted. Before proposing ESDP system we have made investigation throughout different data mining based API recommendation system like MAPO \cite{4}, \cite{5}, GROUMINER \cite{22}, \cite{23} and MACs \cite{9}. Sources are taken from open source repositories in MAPO. They have not used any other sources like regular update, external APIs, standard libraries as their sources of repository. GROUMINER \cite{22}, \cite{23} also have built repository from the open source projects available on the internet. But, MACs have mined sources according to the user query instantly given by the user following post mining approach. Before mining they have dynamically built the repository using Koders. 

Some of the works have not considered APIs from the standard library for a particular platform like Java or C as well as APIs found from the regular searches and APIs from the external APIs in the system. 
If the open source projects and the code search engines like Koders.com or Google code search \cite{16} become the only source of building repository then the mined repository will be obsolete and expired after certain period of time. We have five different sources of repository that provides APIs and class sources to form ESDP central repository as shown in Figure 1.

 Considering this phenomenon the usability of the previously completed repository we also have taken sources of some successfully completed projects from a software company \cite{24}. 
As instantaneous searching has some drawbacks so we have collected and stored the Trending Search Terms of a code search engine in the ESDP repository. But if the repository is not updated after certain duration then the mined repository will be obsolete to provide exact match.So initially we update our central within three months of interval. Lastly, ESDP API developer will write and augment newer APIs to survive and sustain with the critically changed API pattern that the programmers encounter. 

\subsection{Source Abstraction for Data Mining}
In data mining heuristic basically two things happen. In the first step the central repository is preprocessed to an XML repository following a special type of XML conversion strategy. Then in the searching step a data mining algorithm is applied on the XML repository to build mined API XML repository. In the central repository the API and class files are stored as .java or .jar files. First we extract these to code readable files. Then we translate the codes to an abstract form. The common item form is expressed as shown in the Table I and II.

In the second real example as shown in the Table I and II, a method invocation method\_A with a parameter type of java.lang.Stringand return void type appearing in the method method\_C() that is inside the class  Class\_B which is under the package com.

In our ESDP tool 17 types of items are considered purposing research evaluation. But for the brevity only a few items are shown in the Figure 2. 

Then we cover the abstraction code with the XML meta tag. An example is given in later section of Figure 4. The transaction represents the field declaration javax.swing.JButton of the class classB in the package pkga and its position. 
\begin{table}[h!]
\caption{Simulation Parameters in Close Look.}

\centering
\begin{tabular}{||p{1cm} | p{2cm}| p{4cm}||} 
 \hline
\textbf{SL} & \textbf{Generic Example} & \textbf{Real Example }\\ 
 \hline\hline
01 & type, name, entity, location & FD, dom.ASTParser, com.Test:05 \\ 
 \hline
02 & type, name, entity, location & MI, method\_A(java.lang.String): void, com.class\_B.method\_C():130  \\
\hline
\end{tabular}
\end{table}

\begin{table}[h!]
\caption{Detail of Source Abstraction.}

\centering
\begin{tabular}{||p{2cm}| p{5cm}||} 
 \hline
\textbf{Key Term} & \textbf{Elaboration} \\
\hline
FD & Implies for the Field Declaration. FD is an item.\\
\hline
dom.ASTParser & The Field Declaration (FD) statement. dom.ASTParser is an item name.\\
\hline
Com.Test:05  & Denotes that the declaration statement appears at the line 5 within class Test, of package com. \\
\hline
\end{tabular}
\end{table}
A transaction is the set of entities simultaneously used in a block such as class block or method block. ESDP recommends the sequential API code snippets and each recommendation includes several statements. The amount of statements in a sequence is called k-sequence. We took the product of k and the support value of the sequence to the rank the API pattern in the mined XML repository. We have used a recently proposed sequential pattern mining algorithm, called prefixSpan (Prefix-Projected Sequential Pattern Mining) \cite{14}. The algorithm follows the pattern growth method that does not require candidate generation. It mines the complete set of patterns, but greatly reduces the effort of candidate generation and also reduces the projected repository size and lead to the efficient processing.
\begin{figure}[h]
\centering
\includegraphics[height=4.5in]{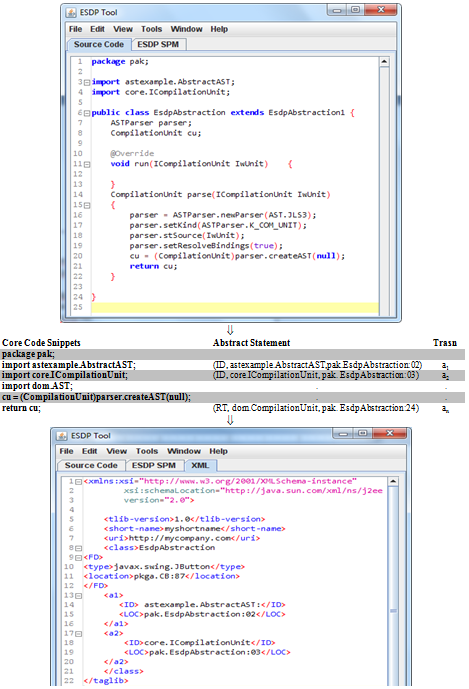}
\caption{Enhancing Software Development Process (ESDP) Framework.}
\end{figure}

\subsection{Searching and Recommendation}
In searching and Recommendation heuristics there are several steps need to be traversed to get a code the skeleton. After the mined XML repository is built, searching is kind of easy with a particular user query statement. Here we get a set of fragment code into a method block after getting searched.  The code snippets are stored in the Mined API XML repository according to their frequency, ranking, support, confidence and with necessary methods and fields’ snippets as shown in the Figure 4. To look for a suggestion developers have to type user query to find the required matching in the Mined API XML Query. It is found by querying the sequential pattern rules with the statement. An example of user query that the user is writing a class called SearchTest is shown in the Figure 3.
\begin{figure}[h]
\centering
\includegraphics[width=3.2in]{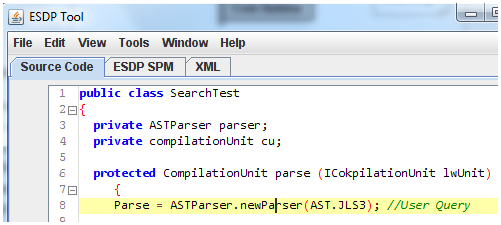}
\caption{ A user query is marked when the developer is writing codes}
\end{figure}

While he is typing the line as user query he wants suggestions from the ESDP repository. The certain statement he marks as user query will be sent to the Mined API XML Repository to find the match. Because the statements and the associated methods and attributes are already stored in the Mined API XML Repository with their support, confidence, rank and sequence number. 

We use it as the input to query the relevant statement sequences. The example shows that several statements sequences are ranked with their scores as shown in the Figure 5. If the developer choose the first match from the given recommendation then the associated code snippets belongs to the background of that match will be retrieved from the XML repository.
\begin{figure}[h]
\centering
\includegraphics[width=3.2in]{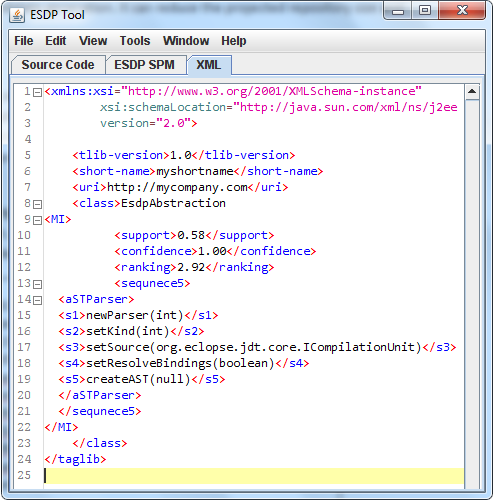}
\caption{ Match is stored in Mined API XML Repository}
\end{figure}

Then the recommended statements are updated and refactored as shown in the preview window of ESDP plug-in inside the extended InetllijIdea IDE
\begin{figure}[h]
\centering
\includegraphics[width=3.2in]{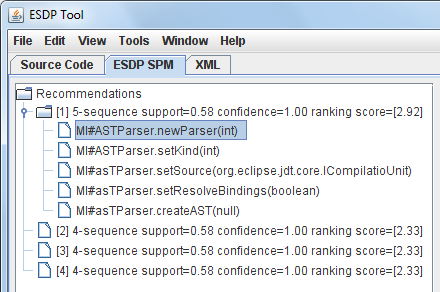}
\caption{ Suggestion after Sequential Pattern Mining in response to the user query
}
\end{figure}

\section{Experimatal Evaluation}
 To evaluate performance and effectiveness of the ESDP repository system several open source projects have been applied it as shown in the Table III. The experiments were carried out in a computer with Windows 7 Operating System, Intel Core I 5 Processor, RAM of 4GB with 3G internet connection of local operator. 

\begin{table}[h!]
\caption{Source Abstraction.}

\centering
\begin{tabular}{||p{.2cm}| p{.6cm}| p{.4cm} | p{.6cm} | p{.6cm} | p{4cm}||} 
 \hline
\textbf{ID} & \textbf{Project} & \textbf{File} & \textbf{Method} & \textbf{USG Patt} & \textbf{Prominent API} 
 \\
\hline
P1 & jEdit 3.0 & 14& 74 & 8 & jmlspecsorg.eclipse.core.runtime.Plugin \\
\hline
P2  & Log4J 1.2.15  & 17 & 79 & 4 & org.apache.commons.codec.binary.Base64\\
\hline 
P3 &  Jigsaw 2.0.5 & 11 & 28 & 7 & Javax.sql\\
\hline 
P4  & Struts 1.2.6  & 13 & 109 & 8 & oracle.core.lmx \\
\hline
P5 &  Fluid VC12.05  & 15 & 106 & 6 &  com.mysql.jdbc \\
\hline
\end{tabular}
\end{table}
The table shows the number of files, methods, usage patterns and the prominent API used in that project. These projects are applied to build our Mined API XML Repository. Then the searching is experimented using user queries on ESDP, MAC and MAPO to check the response time. Here the Table IV and Figure 6 shows the comparisons that are the  required time to respond among different tools. We can see ESDP takes quite lesser time than others.

\begin{table}[h!]
\caption{ Source Abstraction.}

\centering
\begin{tabular}{||p{.2cm}| p{2cm}| p{1cm} |c|c||} 
 \hline
\textbf{UQ} & \textbf{Search Terms} & \textbf{MAC[9]} &\textbf{ MAPO[4,5]} & \textbf{ESDP} \\
\hline
1 & Connection & 0.59 & 5.39 & 0.15 \\
\hline
2  & XMLParser & 0.83 & 6.23 & 0.10\\
\hline 
3 &  getConnection() & 0.55 & 5.56 &  0.09l\\
\hline 
4  & ActionListener  & .74 & 4.96 & 0.10\\
\hline
5 &  InputMissmatch Exception & .67 & 5.35 & 0.17 \\
\hline
\end{tabular}
\end{table}

Five different user queries (UQ) are taken to see the search first and second matching.  When we search for the first user query the result for the first match is found within 6 results by MAC \cite{9}, within 2 results by MAPO \cite{4}, \cite{5}and within 1 result by our ESDP tool.  

The System was manually designed for experiment purpose. Similarly for the other user queries the found results for both first matching and second matching is listed in the following Table V.
 \begin{figure}[h]
\centering
\includegraphics[width=3.5in]{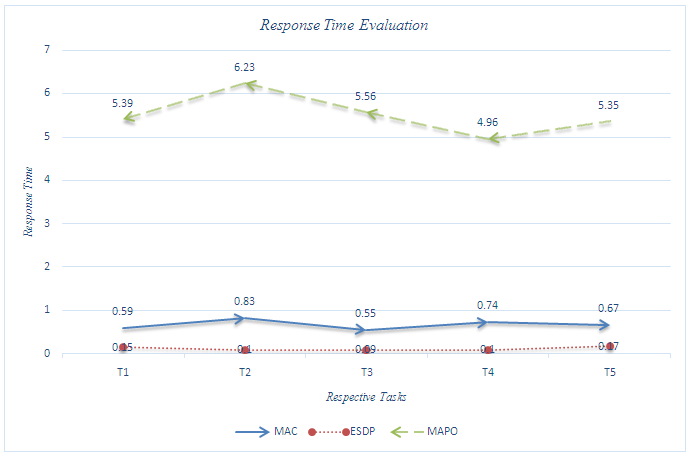}
\caption{A Required Response time comparision among ESDP, MAC and MAPO.
}
\end{figure}

\begin{table}[h!]
\caption{Source Abstraction.}

\centering
\begin{tabular}{||c|c|c|c|c|c|c||} 
 \hline
\textbf{UQ} & \multicolumn{3}{|c|}{\textbf{First Matched}} & \multicolumn{3}{|c|} {\textbf{Second Matched}} \\ [0.5ex] 

\hline
& \textbf{MAC} & \textbf{MAPO} & \textbf{MAC}&\textbf{ ESDP}  & \textbf{ MAPO} & \textbf{ESDP}\\
\hline
UQ1 & 6 & 2 & 1 & 7 & 3 & 2 \\
\hline 
UQ2 & 2 & 1 & 1 & 3 & - & 1\\
\hline  
UQ3 & 2 & 2 & 2 & 2 & 3 & 2 \\
\hline
UQ4 &  - & 1 & 1 & 1 & 2 & 1 \\
\hline
UQ5 & 2 & - & -& - & 2 & 1\\
\hline
\end{tabular}
\end{table}
Table V that our ESDP approach gives less recommendations comparing with the MAC \cite{9} and ESDP tool. 

To see the efficiency with respect to the error vulnerability two different teams were formed to see the empirical user evaluation. The first team was called Experimental Team and second team was called ESDP Team. Both teams have two members. Then they were said to do three different tasks as shown in the Table VI. The experimental team used the manual process using Intellij Idea \cite{15}, \cite{25} IDE and the second team use the IDE where ESDP tool was integrated. 
\begin{table}[h!]
\caption{Source Abstraction.}

\centering
\begin{tabular}{||c|c|c||} 
 \hline
\textbf{Track ID} & \textbf{Description} & \textbf{API Calls} \\ 
\hline
T-001 & Add a context menu to an editor & 5 \\
\hline
T-002 &  Update the name and the bounds of a figure & 4 \\
\hline
T-003 & Save the content of a editor  & 8\\
\hline
\end{tabular}
\end{table}
For the first experiment the Experimental Team consisting of members 1 and 2 brought 6 errors while they do their given tasks where as the ESDP teams faced only 2 errors as per the observation which is shown in Table VII and VIII. 
\begin{table}[h!]
\caption{Source Abstraction.}

\centering
\begin{tabular}{||c|c|c|c|c|c|c||} 
 \hline
\textbf{Tasks} & \multicolumn{3}{|c|}{\textbf{Experimental Team}} &\multicolumn{3}{|c|} {\textbf{ESDP Team}} \\ [0.5ex] 
\hline
& \textbf{M1} & \textbf{M2} &\textbf{ Total}& \textbf{M3}  &  \textbf{M4} & \textbf{Total}\\
\hline
T-001 & 0 & 1 & 1 & 1 & 0 & 1 \\
\hline 
T-002 & 1 & 2 & 3 & 0 & 1 & 1\\
\hline  
T-003 & 0 & 2 & 2 & 0 & 0 & 0 \\
\hline
\multicolumn{3}{|c|}{Grand Total=} & 6 & \multicolumn{2}{|c|}{Grand Total=}& 2\\
\hline
\end{tabular}
\end{table}
For the second empirical study team members were shuffled. That means the member 1 and 2 works for the ESDP team and the rest of the member 3 and 4 worked for the Experiment team. And in that case the given tasks were not changed that they were said to do the same tasks.

From the Empirical Evaluation we see that the ESDP teams face less number of errors than the Experimental teams. So It can be said that ESDP system works better that the usual process. The experimental result is shown in the Table VIII.
\begin{table}[h!]
\caption{Source Abstraction.}

\centering
\begin{tabular}{||c|c|c|c|c|c|c||} 
 \hline
\textbf{Tasks} & \multicolumn{3}{|c|}{\textbf{Experimental Team}} &\multicolumn{3}{|c|} {\textbf{ESDP Team}} \\ [0.5ex] 
\hline
& \textbf{M1} & \textbf{M2} &\textbf{ Total}& \textbf{M3}  &  \textbf{M4} & \textbf{Total}\\
\hline
T-001 & 2 & 2 & 4 & 3 & 2 & 5 \\
\hline 
T-002 & 1 & 3 & 4 & 5 & 3 & 8\\
\hline  
T-003 & 0 & 2 & 2 & 4 & 3 & 7 \\
\hline
\multicolumn{3}{|c|}{Grand Total=} & 10 & \multicolumn{2}{|c|}{Grand Total=}& 20\\
\hline
\end{tabular}
\end{table}
\section{Possible threats to experimetal evaluation}
We have evaluated our tool’s response time, efficiency and error vulnerability with the experimental set up customized and dedicated to our research. So far we have ever seen that the result differs if the system is modified. For that we have marked some threats against the accuracy of our evaluation and experimental study.   

The first threat is that how correctly we have mined our data from the central repository and how correctly we have translated our source abstraction to XML abstraction.  Another thing is that the mining algorithm is an important factor. If the algorithm is changed then our results is also changed.  In some cases it seems to us the existing search engine brings better outcome as they have broader search area and very strong algorithm. 

Our empirical study involves human figures. And we know that the particular programming capabilities of a person as team member may bias the results.  So if the team members are replaced the results may vary. 

The results observed in the empirical study may not be applicable to the programming tasks other than those considered in the study, being a threat to the external validity. If the tasks mentioned out there in the study changes the results may also be changed. 

Before the evaluation the team members were well trained. The receiving capacity of team members might vary. So the of learning curve of these members may affect the results. 
\section{Conclusions}
ESDP repository is more enriched than any other existing approaches. That is why it is able to recommend the sequences more accurately within least interval of time. For data mining we have used prior mining process. That means the source abstraction is mined as a whole using data mining algorithm and then it is stored in the XML data storage. For this reason it is able to work in the client system without being interfered by any server dependency. In response to the user query it provides recommendation sequences of usage pattern with necessary code skeleton in background. Then the code skeleton is refactored and updated with the snippets it needs. ESPD concept is different from most other approaches mentioned in many extents. In future we intend to work with our ESDP system purposing professional implementation after eliminating the flaws that are found in our current research.

\bibliographystyle{IEEEtran}

\bibliography{IEEEabrv}

\end{document}